# Local Elasticity in Nonlinear Rheology of Interacting Colloidal Glasses Revealed by Neutron Scattering and Rheometry

Zhe Wang,*[a,b,c] Takuya Iwashita,[d] Lionel Porcar,[e] Yangyang Wang,[f] Yun Liu,[g] Luis E. Sánchez-Díaz,[c] Bin Wu,[c] Guan-Rong Huang,[h] Takeshi Egami,[i] Wei-Ren Chen*[c]

[a]Department of Engineering Physics, Tsinghua University, Beijing 100084, China

[b]Key Laboratory of Particle & Radiation Imaging (Tsinghua University), Ministry of Education, Beijing 100084, China

[c]Neutron Scattering Division, Oak Ridge National Laboratory, Oak Ridge, TN 37831, USA

[d]Department of Electrical and Electronic Engineering, Oita University, Oita 870-1192, Japan

[e]Institut Laue-Langevin, B.P. 156, F-38042 Grenoble CEDEX 9, France

[f]Center for Nanophase Materials Sciences, Oak Ridge National Laboratory, Oak Ridge, TN 37831, USA

[g]Center for Neutron Research, National Institute of Standards and Technology, Gaithersburg, MD 20899-6100, USA

[h]Physics Division, National Center for Theoretical Sciences, Hsinchu 30013, Taiwan

[i]Department of Materials Science and Engineering and Department of Physics and Astronomy, The University of Tennessee, Knoxville, TN 37996-1508, USA

E-mail: zwang2017@mail.tsinghua.edu.cn; chenw@ornl.gov

## Abstract

The flow of colloidal suspensions is ubiquitous in nature and industry. Colloidal suspensions exhibit a wide range of rheological behavior, which should be closely related to the microscopic structure of the systems. With *in-situ* small-angle neutron scattering complemented by rheological measurements, we investigated the deformation behavior of a charge-stabilized colloidal glass at particle level undergoing steady shear. A short-lived, localized elastic response at particle level, termed as transient elasticity zone (TEZ), was identified from the neutron spectra. The existence of the TEZ, which could be promoted by the electrostatic interparticle potential, is a signature of deformation heterogeneity: The body of fluids under shear behaves like an elastic solid within the spatial range of TEZ but like fluid outside the TEZ. The size of TEZ shrinks as the shear rate increases in the shear thinning region, which shows that the shear thinning is accompanied by a diminishing deformation heterogeneity. More interestingly, the TEZ is found to be the structural unit that provides the resistance to the imposed shear, as evidenced by the quantitative agreement between the local elastic stress sustained by TEZ and the macroscopic stress from rheological measurements at low and moderate shear rates. Our findings provide an understanding on the nonlinear rheology of interacting colloidal glasses from a micro-mechanical view.

## 1. Introduction

Flowing colloidal suspensions are of great importance in our life as well as in a wide variety of industrial applications, such as pharmaceuticals, polymer processing, cosmetics, and transportation technologies. Therefore, there has been much interest in understanding the flow behaviors of colloids.[1,2] The simplest form of colloidal suspensions is the suspension of hard spheres. Extensive computational,[3,4] theoretical[5-7] and experimental investigations of scattering[8-10] and imaging techniques,[11-13] have been performed to study the rheology of hard-sphere colloids. These results significantly broadened our knowledge on how the microscopic structure and flow of hard-sphere colloids are determined by the volume fraction of the colloidal particles and the shear rate $\dot{\gamma}$. Nevertheless, a large amount of colloidal suspensions of everyday and technological importance are not hard-sphere systems, but are characterized by more complicated interparticle interactions. These interactions, such as the electrostatic repulsion and Van der Waals attraction, extend far beyond the range of the excluded particle volume.[14] Because of the extended range of interaction, their rheological properties are often rather different from those of hard-sphere colloidal suspensions at the same volume fractions.[2,7] The microscopic mechanism of the flow of interacting colloidal suspensions demands further studies.

In this work, we investigate the relation between the microscopic structure and rheology of a charge-stabilized colloidal glass as a model colloidal system with soft repulsive interactions. One reason for the current excitement stems from the description of flow based on the concept of "dynamical heterogeneity" – the spatial inhomogeneity in the relaxation dynamics or local configurational rearrangements.[15,16] For example, Yamamoto and Onuki illustrated how the heterogeneity in bond breakage influences the nonlinear rheology of highly-supercooled liquids.[17,18] Particularly, computer simulations suggest that the shear thinning phenomenon is a consequence of decreasing inhomogeneity of flow due to the increasingly frequent configurational fluctuations.[17-20] In the past several decades, there were extensive theoretical and computational studies on the effects of local plasticity in developing a microscopic description of the flow of amorphous solids.[16,21-26] The localized plastic arrangements and their spatial correlation were experimentally examined in colloidal glasses[27-30] and their connection to the shear banding instability was further investigated.[31,32] On the other hand, the role of local elasticity in determining the rheological behavior of soft matters has also been discussed.[33,34] For many soft glasses, such as microgels, polymeric materials, foams and emulsions, the constitutive particles or molecules are deformable and possess significant elasticity. Therefore, the local elasticity can be clearly identified, and is found to deeply influence rheological behaviors.[35] For example, polymeric molecules exhibit a strong entropic elasticity, which leads to the viscoelastic nature of these materials.[36] For many colloidal suspensions, however, the constitutive particles are too hard to contribute any measurable elasticity at typical flow rates. In this case, the local elasticity should be due to the collective rearrangement of particles under deformation, as suggested by a recent computational study.[37] Therefore, experimental identification of the local elasticity and its structural basis is crucial to understand the rheological behaviors, especially the viscoelasticity and the shear thinning phenomenon, of interacting colloidal glasses.

The aim of this work is to experimentally explore the origin of the nonlinear rheology of interacting colloidal glasses from the perspective of dynamical heterogeneity and local elasticity. Small-angle neutron scattering (SANS) technique is a powerful tool to study the microscopic structure of complex fluids at length scales from 1 to several hundreds of nanometers.[38] It has been largely employed to investigate the structure of sheared colloidal suspensions.[2,8-10,39] The analysis of our SANS data shows that the mechanical response of the charge-stabilized colloidal glass to the imposed shear is localized in a *transient elasticity zone* (TEZ), which could be promoted by the electrostatic interparticle potential. The correlation between the TEZ and the mechanical behavior of the sample is supported by the agreement between the microscopic stress revealed by scattering and the macroscopic stress measured by rheometry at low and moderate shear rates. Moreover, the size of the TEZ is found to shrink as the shear rate increases in the shear thinning region, which demonstrates that the shear thinning is accompanied by a diminishing heterogeneity of flow.

## 2. Experimental

### 2.1 Sample

The charge-stabilized colloidal glass used in this study is composed of charged silica particles suspended in a solvent consisting of a mixture of ethylene glycol and glycerol. The proton to deuterium ratio of the solvent was carefully adjusted to avoid possible multiple neutron scattering.[40] The volume fraction of silica particles is 0.4. At this volume fraction, the sample exhibits an evident nonlinear rheological behavior, which will be seen in the next section. The Kob-Andersen mixture of two kinds of silica particles,[41] with diameter of 120 nm and 80 nm in a number ratio of 4:1, was used to avoid shear-induced crystallization.[42,43] The polydispersities of these two kinds of particles are 5.6% and 5.7%, respectively.

### 2.2 SANS experiment

The Rheo-SANS technique under the Couette geometry was employed to study the microscopic structure of the sheared colloids.[44] Figure 1 (a) shows the schematic representation of the SANS experiment. Three principal directions, the flow direction ($v$, denoted as 1), the velocity gradient direction ($\boldsymbol{\nabla v}$, denoted as 2), and the vorticity direction ($\boldsymbol{\omega} = \boldsymbol{\nabla} \times \boldsymbol{v}$, denoted as 3), are defined based on the direction of the applied shear. Two cross sections of the three-dimensional spectrum, namely, the flow-velocity gradient ($\boldsymbol{v} - \boldsymbol{\nabla v}$ or $1-2$) plane and the flow-vorticity ($\boldsymbol{v} - \boldsymbol{\omega}$ or $1-3$) plane, can be measured, as illustrated in Fig. 1 (a). Figure 1(b) shows the SANS spectra obtained from these two planes for the sheared charge-stabilized colloids. When subjected to steady shear, the scattering profiles present elliptical shapes in both configurations. In neither configuration no noticeable scattering signature of shear-induced ordering, such as layer formation, is observed. A similar development is also observed by our complementary Brownian dynamics simulation.[40] Trajectory analysis suggests that the origin of the intensity variation is the local ordering promoted by the anisotropic density fluctuation, instead of the long-range ordering.

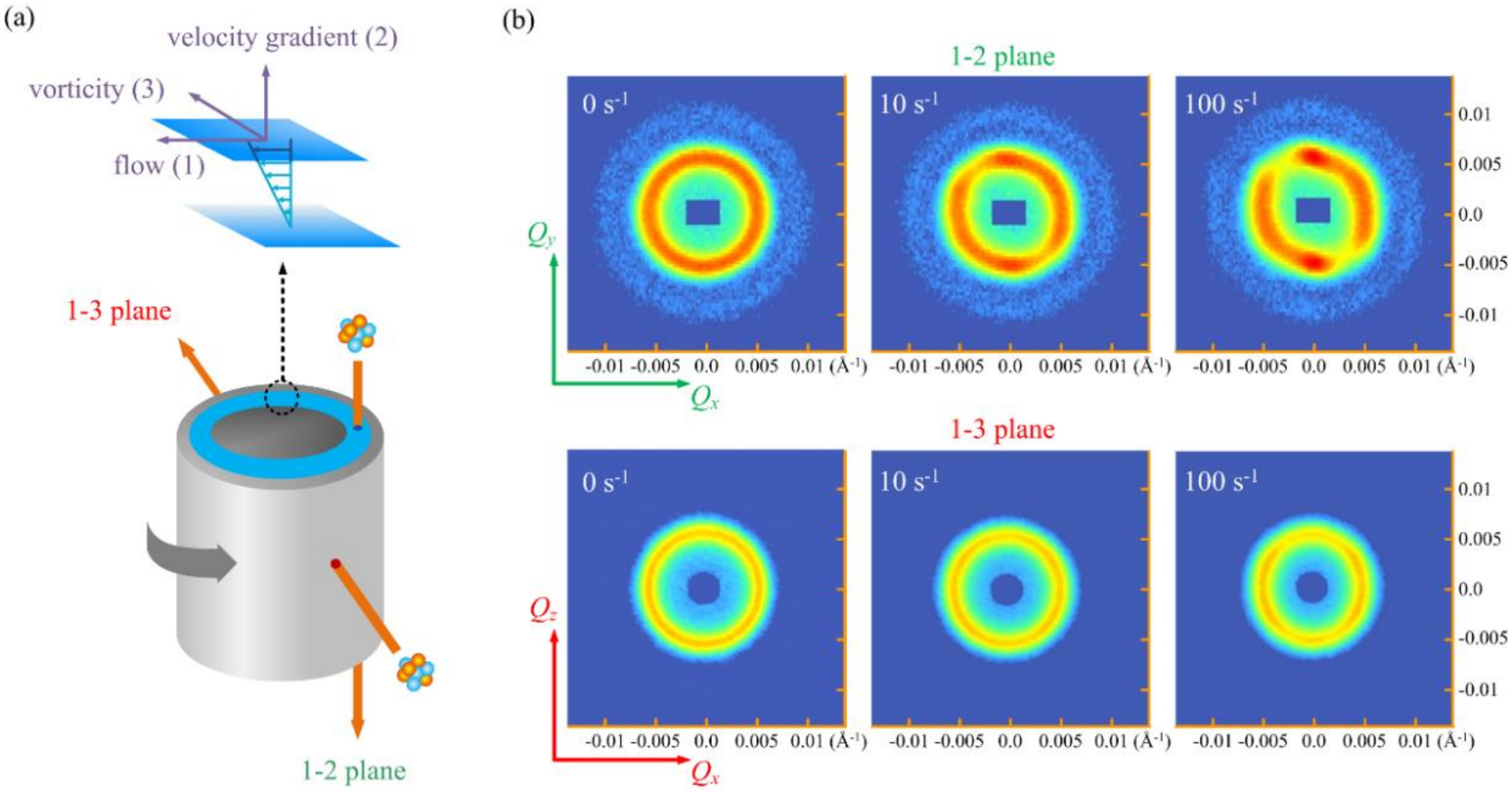

**Fig. 1** (a) Illustration of the Rheo-SANS experiment under Couette geometry. 1, 2, and 3 denote the directions of flow ($\boldsymbol{v}$), velocity gradient ($\boldsymbol{\nabla v}$) and vorticity ($\boldsymbol{\omega}$), respectively. (b) Two-dimensional SANS spectra obtained from the flow-velocity gradient ($\boldsymbol{v} - \boldsymbol{\nabla v}$ or $1-2$) plane and the flow-vorticity ($\boldsymbol{v} - \boldsymbol{\omega}$ or $1-3$) plane at $\dot{\gamma} = 0{,}10$ and $100$ s$^{-1}$ for the charge-stabilized colloidal suspension.

## 3. Results & discussion

### 3.1 Rheological measurements

The small-amplitude oscillatory shear measurement and the steady shear measurement on the charge-stabilized colloidal suspension have been done and the results are shown in Fig. 2. In the linear viscoelastic regime shown in Fig. 2 (a), the dynamic moduli indicate that the sample is an elastic solid in the quiescent state. Results of steady shear measurements given in Fig. 2 (b) show that the sample exhibits a dramatic shear thinning. In following parts, we will provide an illustration based on the cooperative rearrangement of particles under shear.

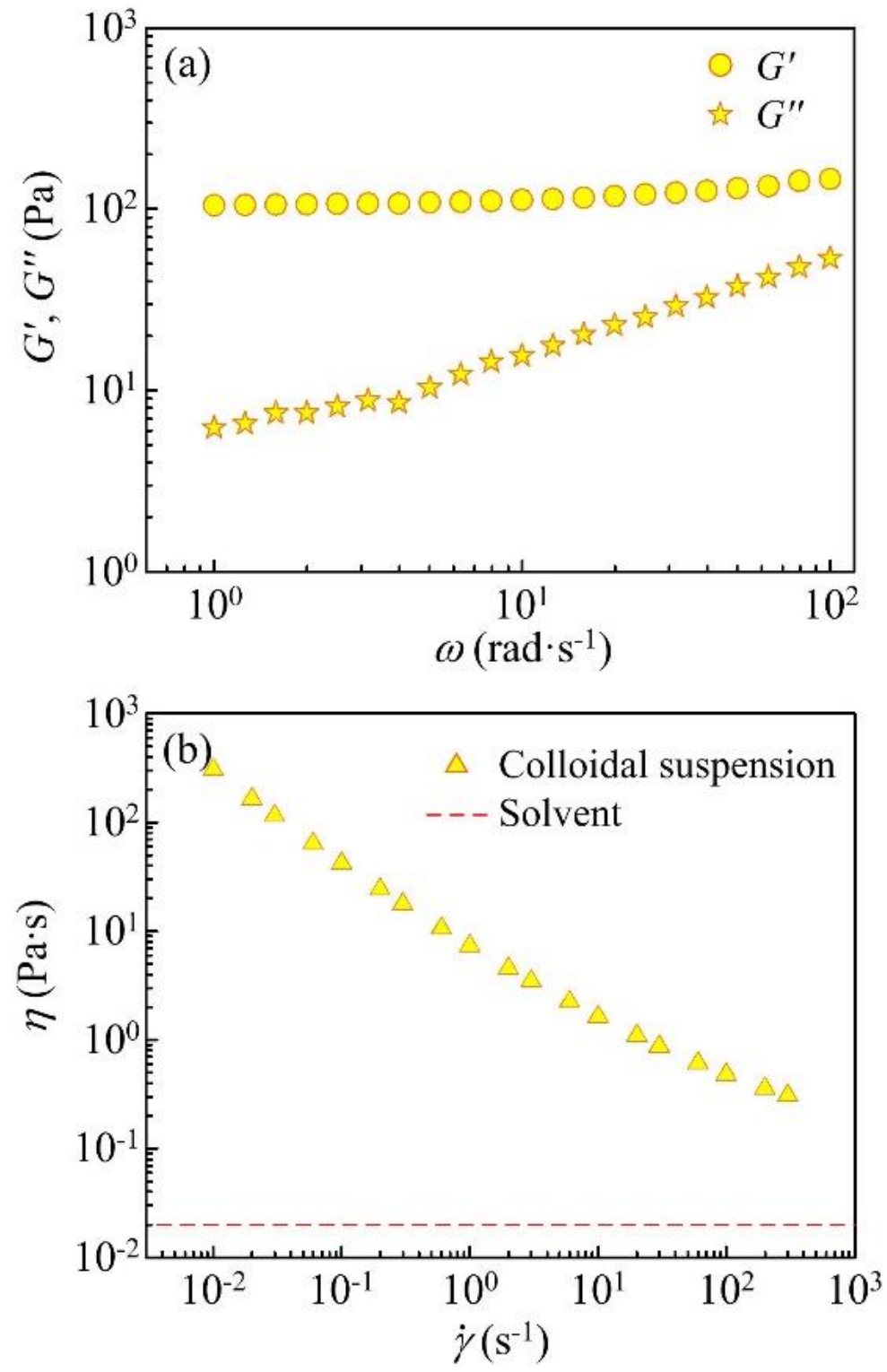

**Fig. 2** Rheological measurements of the charge-stabilized colloidal suspension. (a) Frequency dependence of the storage and loss moduli $G'$ and $G''$. (b) Shear viscosity $\eta$ as a function of shear rate $\dot{\gamma}$.

### 3.2 SANS results

To address the connection between the spatial correlation functions and the flow behavior of the fluids, we adopt a spherical harmonic expansion (SHE) approach for the SANS data analysis. The pair distribution function (PDF) $g(\boldsymbol{r})$ of a sheared fluid can be expressed by SHE as:[45-52]

$$g(\boldsymbol{r}) = \sum_{l,m} g_l^m(r) Y_l^m\left(\frac{\boldsymbol{r}}{r}\right), \tag{1}$$

where $Y_l^m(\boldsymbol{\Omega})$ are the tesseral (real basis) spherical harmonic functions and $g_l^m(r)$ are the expansion coefficients. $g_l^m(r)$ can be determined from SANS experiments by expanding the structure factor, $S(\boldsymbol{Q})$, as:

$$S(\boldsymbol{Q}) = \sum_{l,m} S_l^m(Q) Y_l^m\left(\frac{\boldsymbol{Q}}{Q}\right), \tag{2}$$

which allows us to transform the reciprocal space structural coefficients $S_l^m(Q)$ to the real space coefficients $g_l^m(r)$ using the spherical Bessel transformation:[53]

$$g_l^m(r) = \frac{\mathrm{i}^l}{2\pi^2 \rho} \int S_l^m(Q) J_l(Qr) Q^2 \mathrm{d}Q, \tag{3}$$

where $\rho$ is the number density of the particles and $J_l(x)$ is the spherical Bessel function. Due to the symmetry imposed by shear, $g_2^{-2}(r)$ is the most relevant coefficient that connects the shear-induced structural distortion to the macroscopic properties.[9,54,55] The way of obtaining $g_2^{-2}(r)$ from SANS spectra can be found in Appendix, more details are included in Supplementary Information. For an elastic solid undergoing an affine deformation, $g_2^{-2}(r)$ is proportional to the derivative of the quiescent PDF $g(r)$ when the shear strain $\gamma$ is sufficiently small.[56,57] Namely,

$$g_2^{-2}(r) = -\frac{\gamma}{\sqrt{15}} r \frac{\mathrm{d}g(r)}{\mathrm{d}r}. \tag{4}$$

In Figs. 3(a) to (d) we plot both of $g_2^{-2}(r)$ and $-r\mathrm{d}g(r)/\mathrm{d}r$ determined from the SANS experiment at $\dot{\gamma} = 3, 10, 30$ and $100$ $\mathrm{s}^{-1}$, respectively. It is seen that the characteristic variations of these two functions are generally in phase within the shear thinning regime, which qualitatively agrees with the prediction of Eq. 4. This observation suggests that the system is essentially elastically deformed at these shear rates, even when the system is flowing. Such deformation coherency is also observed in a simulation study on a model metallic liquid.[37]

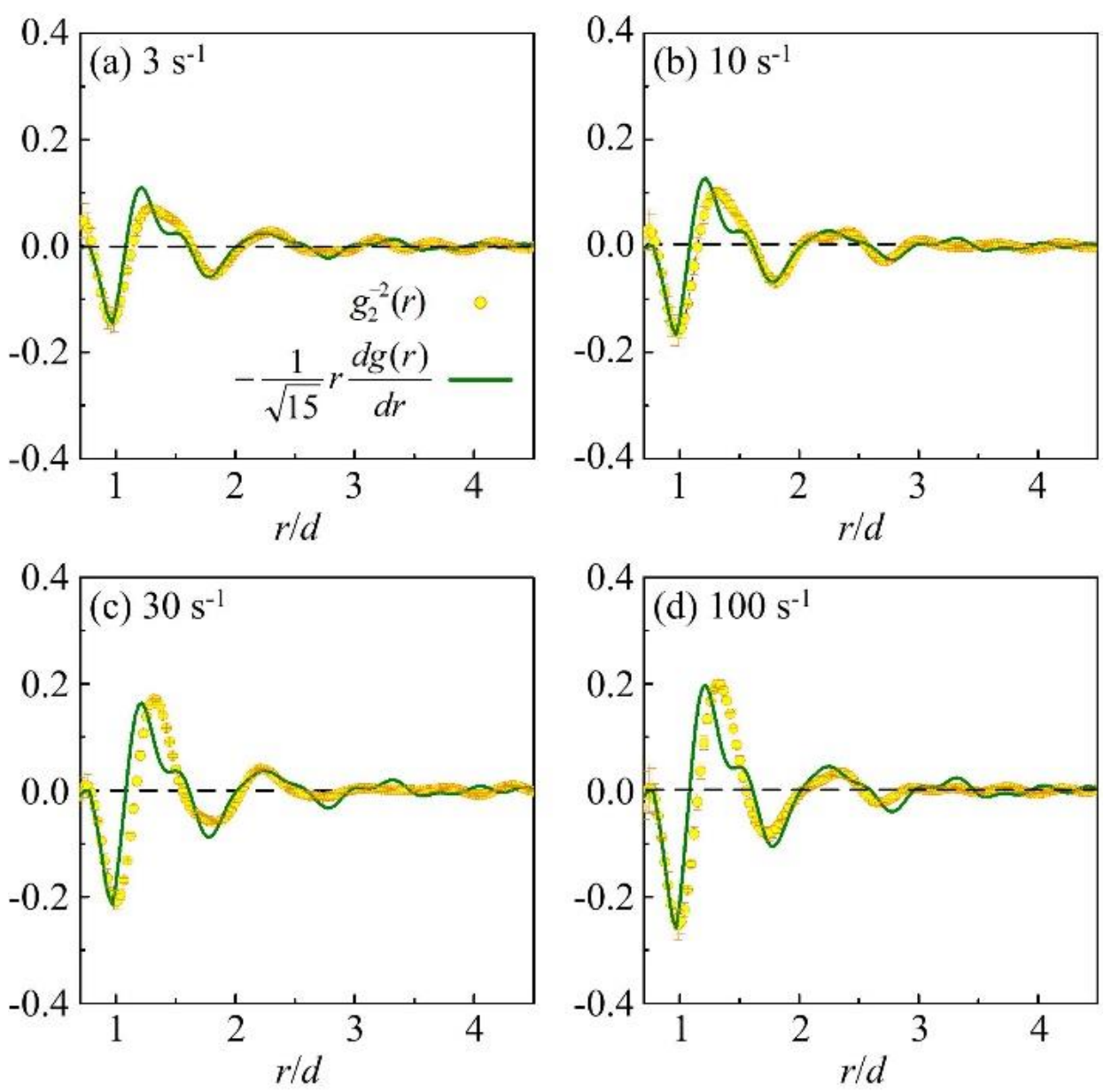

**Fig. 3** Comparison between $g_2^{-2}(r)$ (circles) and $-r\mathrm{d}g(r)/\mathrm{d}r$ (lines). (a)-(d) display the results at $\dot{\gamma} = 3, 10, 30$ and 100 s$^{-1}$, respectively. The magnitude of $-r\mathrm{d}g(r)/\mathrm{d}r$ is scaled to match that of $g_2^{-2}(r)$ for all panels.

In a random stacking of particles, the local configurational environment is known to differ widely from one tagged particle to another. As a result, it is expected that the constant strain picture given by Eq. 4 does not provide a complete description about the microscopic deformation. To further elucidate the structure of the flowing elasticity, we introduce the dependence of $\gamma$ on the spatial range over which the elastic deformation is sustained:[37]

$$\gamma(r) \equiv -\sqrt{15}\, g_2^{-2}(r) / \left[ r \frac{\mathrm{d}g(r)}{\mathrm{d}r} \right], \tag{5}$$

In Figs. 4(a) to (d) we present the γ(r) for the charge-stabilized colloidal suspension at $\dot{\gamma}$=3,10,30 and 100 s-1, respectively. We would like to point out that the extraction of γ(r) does not involve any model fitting but only Bessel transforms and data binning. A region of effectively nonzero γ(r) with a spatial range of several particle diameter d is observed for all measured $\dot{\gamma}$. We name this region transient elasticity zone (TEZ): Within the spatial range of this region, the local structure undergoes an elastic deformation with an average strain given by $\gamma(r)$ when the system is under steady shear. Beyond this region, the particle motion is dominated by liquid-like random displacements. This localized elastic response survives only for a certain lifetime before it relaxes by flow and diffusion. The existence of TEZ suggests the dynamical heterogeneity in the mechanical response of the system to applied shear. $\gamma(r)$ can be considered as a correlation function that describes a cooperative region characterized by mechanical coherency in the flow. Based on the previous simulation study,[37] a Gaussian function is used to model the landscape of $\gamma(r)$:

$$\gamma(r) \approx \gamma_M \exp\left[-\frac{(r-p)^2}{2\delta_{\mathrm{TEZ}}^2}\right], \tag{6}$$

where $p$ is the peak position, $\delta_{\mathrm{TEZ}}$ is the standard deviation of the Gaussian distribution, and $\gamma_M$ is the average maximum strain of the TEZ. The fit curves with Eq. 6 are also shown in Figs. 4(a) to (d). Accordingly, a specific length scale $\xi_{\mathrm{TEZ}} = p + \sqrt{2\ln 2}\delta_{\mathrm{TEZ}}$ is defined to represent the correlation length of the cooperatively elastic deformation in the steady flow. It is seen that $2\xi_{\mathrm{TEZ}}$ denotes the "full width at half maximum" of the TEZ, and can be considered as the size of TEZ. Its shear-rate dependence is shown in Fig. 4(e).[58] A decrease of elastic coherency is revealed by the shrink of the TEZ size from about $6d$ to $4d$ as $\dot{\gamma}$ increases from 1 to $300$ s$^{-1}$. Meanwhile, an increase in $\gamma_M$ from approximately 0.045 to 0.11 is also revealed.

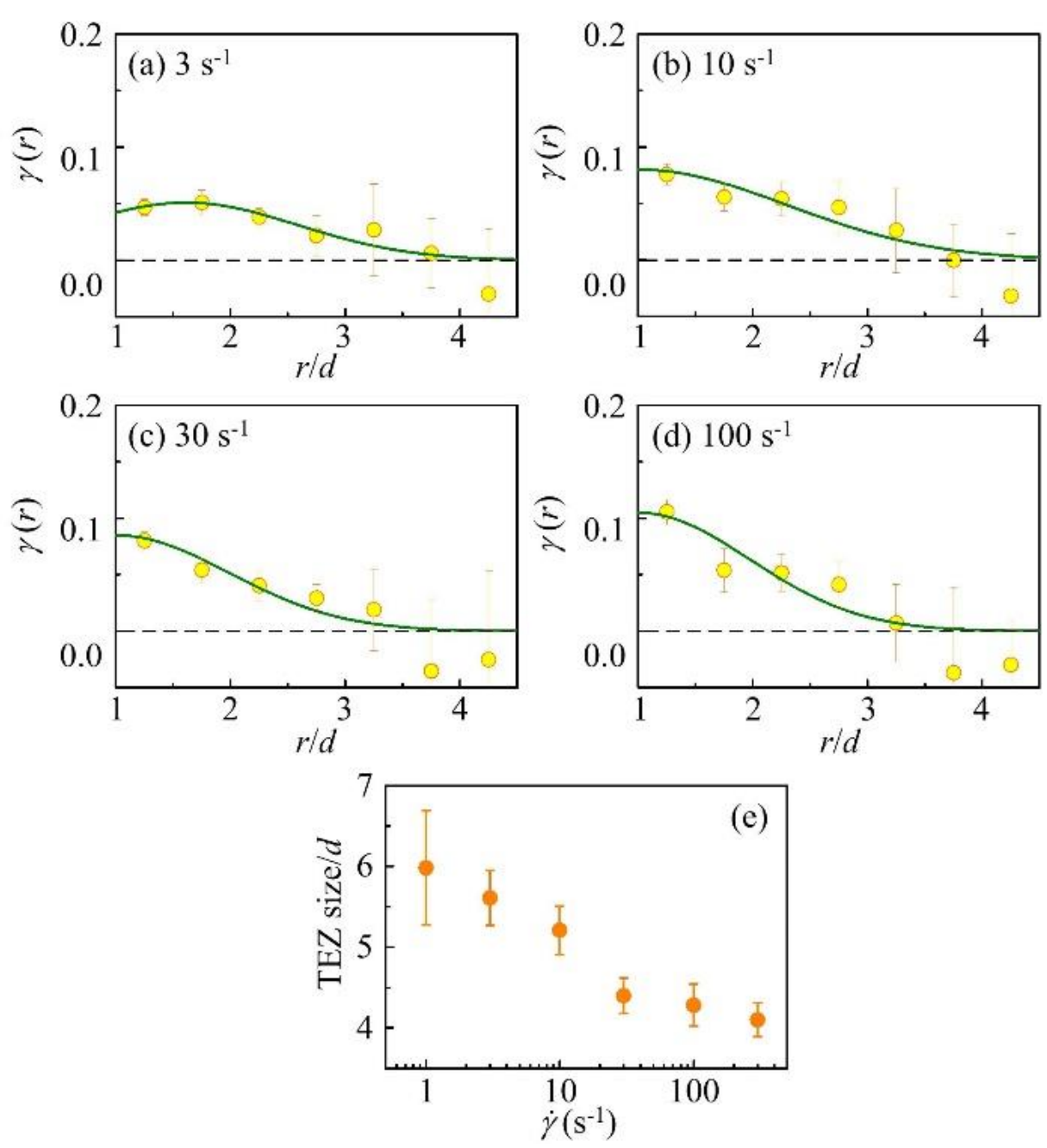

**Fig. 4** (a)-(d): $\gamma(r)$ determined by Eq. 5 at $\dot{\gamma} = 3, 10, 30$ and 100 s$^{-1}$, respectively (yellow circles). The solid curves denote the fitting results by the Gaussian function Eq. 6. Panel (e) summarizes the size of the TEZ $2\xi_{\mathrm{TEZ}}$ as a function of $\dot{\gamma}$.

From Fig. 4 it is seen that at $r > \sim 3d$, the values of $\gamma(r)$ are with large uncertainties and the dependence of $\gamma(r)$ on $r$ becomes irregular. Numerically this is due to the division between two small numbers. As evidenced by Fig. 3, at $r > \sim 3d$, both $g_2^{-2}(r)$ and $r\mathrm{d}g(r)/\mathrm{d}r$ are small, which suggest the loss of structural order and correlation at far distances. In this case, the local elastic coherency is no longer significant.

**3.3 Discussion**

The above analysis reveals a micro-mechanical picture for the deformation of the charge-stabilized colloidal glass. In this system, the changes of the momentum and position of a particle can instantaneously influence surrounding particles through the extended-range electrostatic interaction. Consequently, the particles within a certain spatial range undergo elastic coherent deformation in response to the imposed shear. During this process, a reference particle retains its original neighbors until the stress generated by shear is sufficient to cause local configurational rearrangement. The deformation and yielding of the TEZ are ubiquitous and persistently successive at the particle level. Note that, there should be a structural unit that can store and release elastic energy in viscoelastic materials. Thus, the observation of TEZ is conceptually important for understanding the strong viscoelasticity exhibited by the charge-stabilized colloidal suspension.[59]

In the above mechanism, the interparticle electrostatic repulsion acts like a free energy barrier to resist the applied strain. Therefore, it is crucial in forming the local elasticity in the flow of charge-stabilized colloids. In fact, at the volume fraction of 0.4, it is known that the hard-sphere colloidal suspension is highly fluid, which suggests the absence of TEZ. The hard-sphere suspension exhibits evident elasticity only when the volume fraction is higher than about 0.58, in which case the excluded volume effect is significant.[7]

Having established the picture of transient local elasticity from SANS experiment, we now proceed further to explore the role of the TEZ in the nonlinear rheological behavior of interacting colloidal suspensions. In the flowing charge-stabilized colloids, the elastic stress sustained by TEZ is estimated as $\sigma_{\mathrm{TEZ}} = G'\gamma_M$, where $G'$ is the modulus of the local elasticity that is similar to the storage modulus given in Fig. 2(a).[60] As given in Fig. 5(a), the microscopically determined stress $\sigma_{\mathrm{TEZ}}$ is seen to be in a quantitative agreement with the macroscopic shear stress $\sigma_{\mathrm{CC}}$ ($\sigma_{\mathrm{CC}} = \eta\dot{\gamma}$) determined from rheometry when $\dot{\gamma} \leq 10$ $\mathrm{s}^{-1}$. This agreement is remarkable considering that the two approaches of measuring stress are completely different. It clearly reveals that the elasticity of the TEZ causes the high shear stress, or equivalently the viscosity, in the flow of the charged-stabilized sample. At higher shear rates ($\dot{\gamma} \gg 10$ $\mathrm{s}^{-1}$), $\sigma_{\mathrm{TEZ}}$ considerably deviates from $\sigma_{\mathrm{CC}}$, manifesting the increasing fluidization.

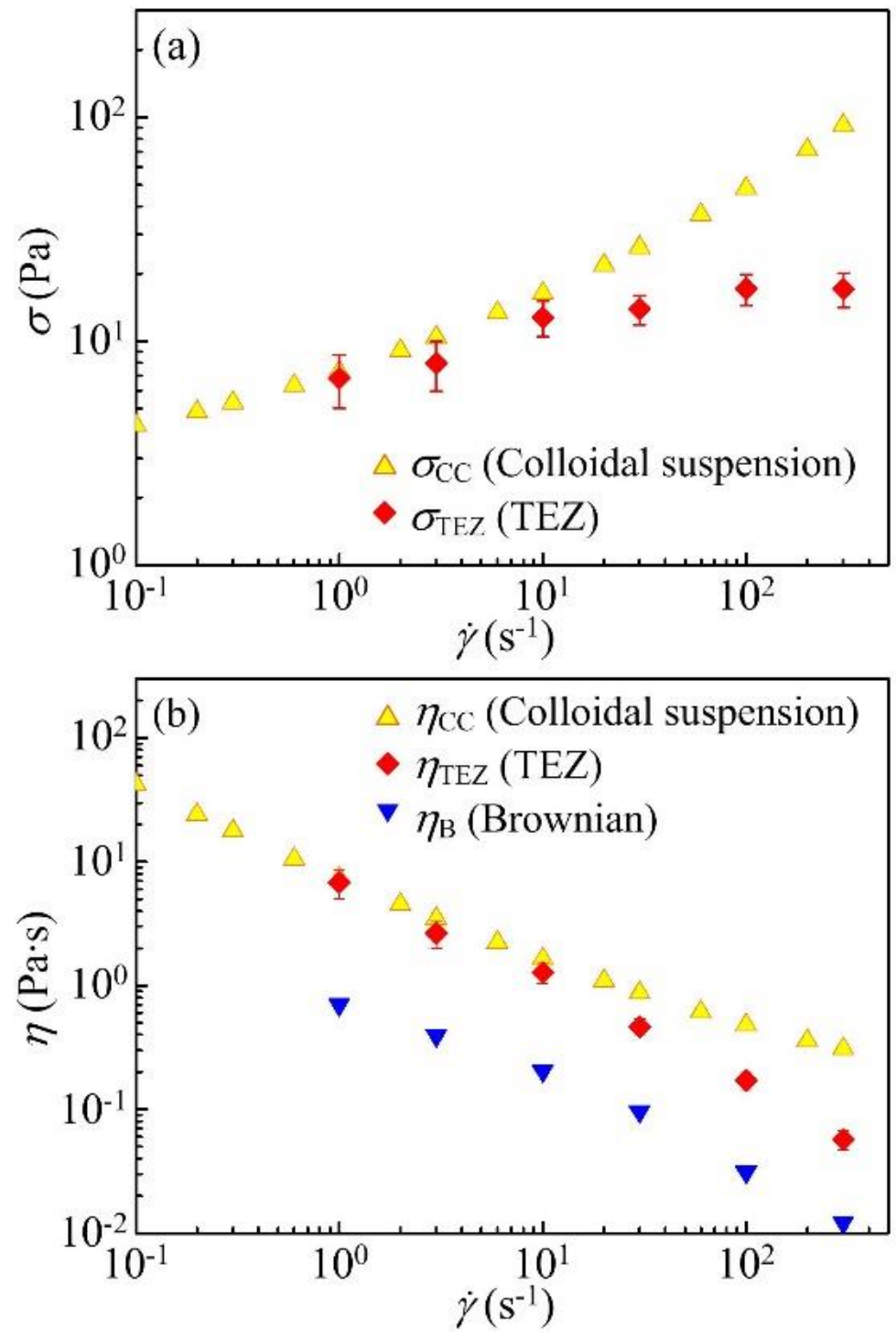

**Fig. 5** (a) Shear stress $\sigma$ and (b) viscosity $\eta$ as a function of the shear rate $\dot{\gamma}$. In panel (a), the elastic stress sustained by the TEZ is also shown. In panel (b), the viscosity contributed by TEZ and the Brownian viscosity contribution are also plotted.

It is known that Brownian and hydrodynamic effects contribute to the viscosity of colloidal suspensions.[4,5] The Brownian viscosity contribution $\eta_{\mathrm{B}}$ can be calculated from $g(\boldsymbol{r})$ [5,61,62] and the result is shown in Fig. 5(b). The hydrodynamic viscosity contribution, estimated with the approximation given in Ref. 61, is found to be well below 0.1 Pa·s. This value is small since the electrostatic repulsion can suppress the hydrodynamic effect by reducing the near-contact lubrication.[61] The viscosity contribution from the TEZ ($\eta_{\mathrm{TEZ}} = \sigma_{\mathrm{TEZ}}/\dot{\gamma}$) is plotted in Fig. 5(b). The viscosity of the sample $\eta_{\mathrm{CC}}$ measured by rheometry is also shown in Fig. 5(b) for comparison. It is seen that for the flowing charge-stabilized colloids, $\eta_{\mathrm{TEZ}}$ is much larger than the Brownian viscosity contribution $\eta_{\mathrm{B}}$. This result agrees with the theoretical prediction that in charge-stabilized colloids the potential viscosity contribution is much stronger than the Brownian viscosity contribution.[61] Summarizing these results, we confirm that TEZ plays a key role in the nonlinear rheology of the sheared charge-stabilized colloids. This is very different from the hard-sphere colloids, in which the shear thinning is mainly attributed to the Brownian effect.[4,11,61]

The concept of dynamical heterogeneity was introduced for the first time to explain the dramatic increase of the viscosity in the glass transition of supercooled liquids or the colloidal glass formation.[63-65] Numerous computational and theoretical studies[66,67] and recent experiments[68] have shown that the drastic increase in viscosity is accompanied by the growth of cooperatively rearranging regions in fluids during the supercooling process. Yamamoto and Onuki generalized this thinking to the flowing supercooled and glassy liquids: With computer simulations, they demonstrated that the size of the region characterizing the collective bond breakage events shrinks during the shear thinning.[17,18] Experimentally, we found that the size of the TEZ decreases with increasing the shear rate (Fig. 4(e)), implying the observed shear thinning behavior is accompanied by diminishing deformation heterogeneity. This result is consistent with the previous computational investigations.

## 4. Conclusions

In summary, using SANS and rheometry, we identify a transient elasticity zone (TEZ) in a charge-stabilized colloidal glass undergoing steady shear. This TEZ, which manifests the local elasticity in flow, is a many-body effect sustained by the electrostatic interparticle repulsion. We show that the TEZ acts as the micro-structural unit that resists the shear from the agreement between the microscopic elastic stress determined by SANS and the macroscopic stress measured by rheometry at low and moderate shear rates. The spatial range of TEZ spans over a distance of a few particle diameters. It is found to shrink with increasing the shear rate, suggesting that the shear thinning is associated with the weakening of the dynamical heterogeneity in the flow. Our findings shed new light on understanding the nature of nonlinear rheology and viscoelasticity of interacting glasses and highly supercooled liquids.

**Appendix**

In this part, we will give a description of obtaining $g_2^{-2}(r)$ from the SANS spectra. First of all, we need to obtain $S_2^{-2}(Q)$ from the spectra. The structure factor of the sheared colloids, $S(\boldsymbol{Q})$, can be expanded by spherical harmonic functions:

$$S(\boldsymbol{Q}) = \sum_{l=0}^{\infty} \sum_{m=-l}^{l} S_l^m(Q) Y_l^m\left(\frac{\boldsymbol{Q}}{Q}\right), \tag{A1}$$

where $Y_l^m(\boldsymbol{\Omega})$ are the real basis spherical harmonic functions defined as:

$$Y_l^m(\boldsymbol{\Omega}) = Y_l^m(\theta, \phi) = \begin{cases} \sqrt{2}\sqrt{(2l+1)\frac{(l-|m|)!}{(l+|m|)!}} P_l^{|m|}(\cos\theta)\sin(|m|\phi) \quad (m < 0) \\ \sqrt{2l+1} P_l^0(\cos\theta) \quad (m = 0) \\ \sqrt{2}\sqrt{(2l+1)\frac{(l-m)!}{(l+m)!}} P_l^m(\cos\theta)\cos(m\phi) \quad (m > 0) \end{cases} \tag{A2}$$

where $P_l^m(x)$ are the associated Legendre polynomials, $\theta$ is the angle with respect to the 3 axis, $\phi$ is the polar angle with respect to the 1 axis. In three-dimensional space the spherical harmonic functions are mutually orthogonal:

$$\int \mathrm{d}\boldsymbol{\Omega} Y_l^m(\boldsymbol{\Omega}) Y_{l'}^{m'}(\boldsymbol{\Omega}) = 4\pi \delta_{ll'} \delta_{mm'}. \tag{A3}$$

Each function has a well-defined parity:

$$Y_l^m(\theta, \phi) \rightarrow Y_l^m(\pi - \theta, \pi + \phi) = (-1)^l Y_l^m(\theta, \phi). \tag{A4}$$

Within the accessed range of shear rate, the contribution from terms with $l = 3, \ldots, \infty$ to spectra should be small compared with the several leading terms. This approximation can be justified from the fact that the measured strain of the TEZ is only around 0.1 or even less. Therefore, Eq. A1 can be approximated as:

$$S(\boldsymbol{Q}) \approx \sum_{l=0}^{2} \sum_{m=-l}^{l} S_l^m(Q) Y_l^m\left(\frac{\boldsymbol{Q}}{Q}\right). \tag{A5}$$

Under the geometry of shear flow, it is seen that $S(\boldsymbol{Q})$ satisfies the conditions $S(Q,\theta,\phi) = S(Q,\pi-\theta,\phi)$ and $S(Q,\theta,\phi) = S(Q,\theta,\phi+\pi)$. These conditions can be justified from Fig. 1(b). Therefore, only terms with even $l$ and $m$ survive, which leads to the following expression:

$$S(\boldsymbol{Q}) \approx S_0^0(Q)Y_0^0\left(\frac{\boldsymbol{Q}}{Q}\right) + S_2^{-2}(Q)Y_2^{-2}\left(\frac{\boldsymbol{Q}}{Q}\right) + S_2^0(Q)Y_2^0\left(\frac{\boldsymbol{Q}}{Q}\right) + S_2^2(Q)Y_2^2\left(\frac{\boldsymbol{Q}}{Q}\right). \quad \text{(A6)}$$

In the $1-2$ plane, we can define the following quantity:

$$S_{2,-2}^{xy}(Q) = \frac{1}{2\pi}\int_0^{2\pi} S\left(Q,\theta=\frac{\pi}{2},\phi\right) Y_2^{-2}\left(\theta=\frac{\pi}{2},\phi\right)\mathrm{d}\phi$$

$$= \frac{1}{2\pi}\int_0^{2\pi} S\left(Q,\theta=\frac{\pi}{2},\phi\right)\frac{\sqrt{15}}{2}\sin(2\phi)\mathrm{d}\phi. \quad \text{(A7)}$$

Since $S(Q,\theta=\pi/2,\phi)$ can be measured from the SANS experiment (see Fig. 1(b)), $S_{2,-2}^{xy}(Q)$ can be obtained easily from the measured spectra in the $1-2$ plane. Combining with Eq. A6, it is straightforward to show that:

$$S_2^{-2}(Q) = \frac{8}{15}S_{2,-2}^{xy}(Q), \quad \text{(A8)}$$

With $S_2^{-2}(Q)$, $g_2^{-2}(r)$ can be obtained by spherical Bessel transformation:

$$g_l^m(r) = \frac{\mathrm{i}^l}{2\pi^2\rho}\int S_l^m(Q)J_l(Qr)Q^2\mathrm{d}Q. \quad \text{(A9)}$$

The calculation of $S_0^0(Q)$ from SANS spectra needs the information on the $1-3$ plane. The detail can be found in Supplementary information.

## Conflicts of interest

There are no conflicts to declare.

## Acknowledgements

The work at Oak Ridge National Laboratory was supported by the U.S. Department of Energy, Office of Science, Office of Basic Energy Sciences, Materials Sciences and Engineering Division. The work at Tsinghua University was supported by the Thousand Talents Plan for Young Professionals from the Chinese Government. The research at Spallation Neutron Source of Oak Ridge National Laboratory was sponsored by the Scientific User Facilities Division, Office of Basic Energy Sciences, U.S. Department of Energy. The rheological characterization was carried out at the Center for Nanophase Materials Sciences of Oak Ridge National Laboratory, which is a DOE Office of Science User Facility. We acknowledge the support of the National Institute of Standards and Technology, U.S. Department of Commerce, in providing the neutron research facilities used in this work. Finally, we gratefully appreciate the D22 SANS beamtime from the Institut Laue-Langevin.

## Notes and references